\documentclass[a4paper,conference]{IEEEtran}
\usepackage[left=1.57cm,right=1.57cm,top=0.95cm,bottom=2.54cm]{geometry}

\usepackage{ifpdf}
\ifCLASSINFOpdf
  \usepackage[pdftex]{graphicx}
\else
  \usepackage[dvips]{graphicx}
\fi
\usepackage{pgfplots}
\usepackage{amsmath}
\usepackage{amsthm}
\usepackage{url}
\usepackage{epstopdf}

\usepgfplotslibrary{groupplots,dateplot}
\usetikzlibrary{patterns,shapes.arrows}
\pgfplotsset{compat=newest}
\usepackage{cite}
\usepackage{dsfont}

\usepackage{xcolor}

\newcommand{\figw}{0.99\columnwidth}
\newcommand{\figwa}{0.94\columnwidth}
\newcommand{\figwb}{0.95\columnwidth}
\newcommand{\figwc}{0.64\columnwidth}
\newcommand{\figwd}{\columnwidth}

\begin{document}

\title{Blockage-Peeking Game of Mobile Strategic Nodes in Millimeter Wave Communications}

\author{Leonardo Badia$^*$ and Andrea Bedin$^{*\dag}$ \\
$*$ Dept.\ of Information Engineering (DEI), University of Padova, Italy \\
$\dag$ Bell Labs, Nokia, Espoo, Finland \\
email: leonardo.badia@unipd.it, andrea.bedin.2@studenti.unipd.it
}

\maketitle
\pagestyle{empty}
\thispagestyle{empty}

\begin{abstract}
Given the importance of line-of-sight in mmWave communications, a strategic adversary can harm a transmission by obstructing the receiver, which in turn can react by trying to move around this hurdle. To expand on this point, we study one such scenario from the perspective of game theory, considering a mobile mmWave receiver and an adversary interacting strategically as players in a zero-sum game, where they want to maximize, or respectively minimize, the spectral efficiency of the communication. To do so, the adversary attempts at screening the receiver's line of sight as an obstacle, while the receiver can move around so as to avoid the blockage. We consider pre-set distances and the choices available to the players are to change their angular coordinates to go around each other. This is framed as a static game of complete information, for which we numerically find the Nash equilibrium in mixed strategies, drawing some interesting conclusions such as connecting it with the beamforming pattern of the transmitter.
\end{abstract}

\begin{IEEEkeywords}
Wireless communications; mmWave; Physical layer security; Game theory; Zero-sum games.
\end{IEEEkeywords}

\section{Introduction}
Physical layer security is a key challenge for wireless communications, because the radio channel is inherently broadcast and its ease of access makes it vulnerable to attacks \cite{liu2016physical}. 
This is relevant to mmWave systems, which are designed to meet demanding throughput, latency, and reliability requirements, but at the same time exploit a physical channel with high losses, directivity, and sensitivity to blockage \cite{niu2015survey}.

A frequent setup for physical layer security in the literature involves the interaction of a legitimate transmitter/receiver system with an adversary whose intent is to disrupt its operations.
This is often explored in the context of jamming~\cite{commander2007wireless,altman2007jamming,scalabrin2015zero,vadori2015jamming}, albeit there may be variations, depending on whether a jammer is just raising the noise floor and thereby choking the Shannon's capacity of the channel, or is also able to perform advanced intrusions such as eavesdropping or spoofing~\cite{manshaei2013game}. Moreover, the ``red/blue'' roles can be reversed in the case of \emph{friendly} jamming \cite{badia2019game}, where jamming is actually the legitimate network's action to prevent attackers from malicious communication exchanges; as a result, this approach can be generalized to other adversarial contexts.

The analysis of this interaction is often done from the perspective of game theory, and specifically framing it as a zero-sum game~\cite{dasilva2011game}, with the value taken as the capacity or the spectral efficiency of the channel, which in turn depends on the signal-to-noise ratio at the receiver's end and therefore on power control and propagation aspects. The players are the legitimate transmitter or receiver (according to where the strategic intelligence is located), acting as a maximizing player, and, opposite to it, an adversary being the minimizer. 

This basic framework can also be expanded, for example, towards higher layers in the protocol stack, whenever a Nash equilibrium (NE) in mixed strategies is interpreted as a pattern of channel access probabilities \cite{ghazvini2012game}, or to account for other characteristics of the player through Bayesian games, e.g., uncertainty about their malicious intent \cite{garnaev2020jamming} or time-varying channel conditions~\cite{sagduyu2011jamming}. 

With respect to the vast existing literature of physical layer security games, where the strategic choices relate to power control or channel allocation \cite{prospero2021resource,altman2007jamming,manshaei2013game,gao2018game}, considerably less attention is paid to the issue of mobility of the involved nodes, which is interesting on multiple levels for mmWave communications.

First of all, mmWave communications can provide high data rates, but when the positions of the users vary, the channel tends to change significantly, depending on whether a line-of-sight (LoS) exists between the transmitter and the receiver \cite{lim2021deep,el2020rapid,korpi2020visual}. At the same time, moving away from a source of noise or changing location to avoid channel fades is a quite natural behavior adopted by real receivers, even those acting without any pretence of optimality \cite{perin2021adversarial,hussain2020mobility}; thus, in a game theoretic context, it can be considered as  available to a rational player \cite{misra2020m}. 
Finally, we point out that, in mmWave communications, we do not require an advanced adversary to undertake elaborate actions, we can focus on the case where it is merely placing an obstacle to cause blockage to the receiver, which in turn tries to evade it \cite{perin2021reinforcement}.

The reaction to an LoS blockage ought to be to not just evade but also foresee the system evolution, especially the possible reaction of the obstacle. To cite a famous example from the Italian literature, albeit unrelated to mmWave communications, Dante Alighieri describes in one of his early poetry \cite{vitanova} how he was exchanging loving looks during Mass with his beloved, which was the most daring way to flirt back then. However, she was married to another man, so the presence of a ``screen woman'' (i.e., a beautiful lady who happened to be in between them) gave him the opportunity to dissimulate his glances towards the real target. Only, his plan backfired when the screen woman believed these looks of affection were really directed to her.

In this paper, we want to address a simpler scenario of physical layer security, and to this end we formalize a ``Blockage-Peeking'' game for mmWave communications, played by two mobile elements, the former being a mmWave receiver and the latter being an adversary that is trying to obstruct the receiver's LoS. Both are modeled as strategic players in a zero-sum setup, which, in spite of being simply formulated as a static game of complete information, is able to provide interesting results, when a realistic model of the surrounding propagation scenario is accounted for.

The paper is therefore giving the following original contributions: first and foremost, up to our knowledge, the idea of a Blockage-Peeking game itself has never been presented before in the literature. The difference with the many studies pertaining to mmWave communications where beamforming or movements of the terminals are exploited to find a clear LoS is that we explicitly consider a strategic setup, where blockage is intentionally caused by a malicious user, and both the receiver and the adversary are rationally able to react.

For the sake of simplicity, we limit the interactions of the players to just the movement around each other. We are aware that the interactions in mmWave transmissions also offer other options to counteract obstacles and bad propagation, whenever they are spontaneously present or caused by malicious attackers, including beamsteering or alignment over different transmission paths \cite{bonjour2015ultra,abdelnabi2020outage}, not to mention the possible use of reflective intelligent surfaces \cite{wang2021joint}. However, in this paper we specifically focus on the issue of movement of the involved terminals, leaving them for future extensions of the present work.

Moreover, as the specific investigation of this paper, we analyze the interaction of the players as a static zero-sum game of complete information, which is the simplest way to introduce a game theoretic rationale. Yet, even this aspect can be seen as a cornerstone for more complex investigations, where the game is made dynamic, or Bayesian elements, i.e., incompleteness of the information available to the players, are inserted \cite{quer2013inter,etesami2019dynamic,gindullina2017asymmetry}.

We analyze the NE, which is found to be unique and in mixed strategy, and we compute it for different propagation setups. In particular, we consider a receiver at a fixed distance from the transmitter, nevertheless being able to move \emph{around} it, which allows to avoid blockage at the price of obtaining a less convenient beamforming pattern of the antennas, and an adversary placing an obstacle in between them, also at a variable angular coordinate. Thus, we perform different investigations varying the distance between the obstacle and the transmitter, and discuss the resulting NE.

While it seems intuitive that the most suitable position for the adversary is the one causing the most severe blockage to the receiver, we found out that the NE gives a more dispersed placement, as the result of a mixed strategy. This can be seen as due to the combination of the primary goal of the attacker to cause havoc, with the general game theoretic principle that a strategic agent should play so that the possible outcomes are all made equivalent for the opponent's payoffs \cite{badia2019game,garnaev2020jamming}. 

For this reason, the position of the obstacle at the NE is a bit slanted towards secondary propagation lobes to prevent the receiver from an easy escape of that blockage. Similarly, the receiver has a mixed strategy that includes the action of aligning with the direct link to the transmitter but also other choices, actually more prominent, following the radiation pattern of the antenna, and the best strategy to maximize the spectral efficiency of the channel is to randomize among them. So, in general, the results are non trivial and they prompt further investigations on this scenario.

The rest of this paper is organized as follows. We review related work in Section \ref{sec:relw}. Section~\ref{sec:mod} describes the underlying models of our analysis, i.e., the mmWave propagation scenario and the resulting strategic interactions. Section~\ref{sec:gam} analyzes it as a static zero-sum game, and discusses its NEs in mixed strategies. Section~\ref{sec:res} presents numerical results and discusses extensions to more advanced game theoretic setups. Finally, the conclusions are drawn in Section~\ref{sec:concs}.

\section{Related Work}
\label{sec:relw}

After many related studies in the last $20$ years, the use of game theory to capture the performance of wireless networks is now a well consolidated topic \cite{dasilva2011game,manshaei2013game}. Game theoretic instruments turn out to be especially valuable whenever the interaction among the network agents is fully distributed and/or there is an underlying scarcity of resource or inherent conflict in the medium access control, as is the case for collision-avoidance techniques \cite{ghazvini2012game}.

However, the inability of accessing the wireless medium, not caused by unfortunate collisions but an intentional action, can also be related to a physical layer security context, as for jamming scenarios \cite{garnaev2020jamming,sagduyu2011jamming,commander2007wireless}.
This is a common example of how multiple wireless agents with contrasting objectives can be framed in a game theoretic setup, and it belongs to a broader class of network security problems, spanning through the multiple layers of the protocol stack \cite{manshaei2013game}.

Many game theoretic formulations consider an adversarial situation that can be framed as a minimax or zero-sum game, possibly with some variations \cite{zhu2010network,badia2019game,moon2015minimax}.
However, while some properties of the physical and medium access layers are often used to identify the strategies available to the players, we believe that the physical positioning of the involved nodes is relatively less explored. Some investigations were performed in \cite{scalabrin2015zero,perin2021reinforcement}, but only related to jamming. In these papers, positioning was considered as influencing the Bayesian types of the players, and subsequently extended to being an active strategic choice of the receiver to evade the jammer.

This concept of dodging a jammer was already explored in \cite{gao2018game} in a game theoretic fashion, but once again from a higher layer perspective, in this case by considering a frequency-hopping scheme.
However, the simple idea of physically moving away from a source of noise or a bad channel fade, despite being part of everyday experience for many personal device users, has not been explored very often. 

For most of the physical-layer related literature, mobility of the terminals is considered harmful to the connection and a source of outage. However, in adversarial scenarios it can also be a way to escape harsh channel conditions caused by a malicious jammer. A handful of papers actually take this into account.
For example, \cite{misra2020m} investigates the mobility of a wireless node to avoid interference, but the scenario is not adversarial, and the interference condition is just incidental.
In \cite{azogu2013new}, the setup is instead explicitly considering jamming, and several strategies are devised to counteract it, one among them being movement of the receiver; however, in this paper the whole area affected is considered to be jammed, and there is no strategic counter-reaction by the adversary. The movement of the users is also addressed in \cite{chrysikos2010wireless} for an adversarial setup, but  to evaluate its impact on the secrecy capacity from an information-theoretic perspective, and it is found that a moving eavesdropper is more harmful.
Finally, \cite{perin2021reinforcement} more directly considers the choice of position as a dynamic game between a receiver and an adversary, but the focus is still on jamming as a zero-sum game.

Nearly no contribution applies this reasoning to the case of mmWave communications, where the issue of mobility is particularly acute. Propagation effects can be altered by mutual movement of the involved objects, and especially the obstruction of LoS is severely affecting mmWave channels \cite{lim2021deep}. For this reason, an interesting game theoretic investigation can focus on a simple behavior by an adversary, which involves no jamming whatsoever, but rather an intentional obstruction of the LoS for the receiver, and the search for a better communication beam. One such investigation was performed in \cite{hussain2020mobility}, but not from a game theoretic perspective, and the blockage that the users can experience is, once again, assumed to be incidental and not intentional.

To sum up, we believe that the present paper fills a gap in the existing literature. We give a game theoretic twist to a scenario that is expected to become extremely common in future generation networks, where mmWave communication will be pervasive. And we combine topics that, despite their importance, are seldom addressed together. Even though game theory is often used for security investigations, it is rarely combined with basic aspects of the physical layer, such as mobility of the nodes and its impact on the propagation scenario, which are particularly relevant.

\begin{figure}[tb]
\hspace{0.3cm}\resizebox{\figwa}{!}{\input{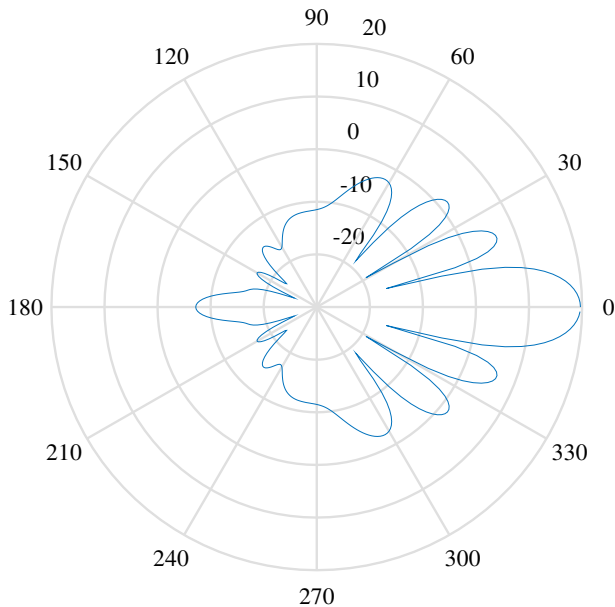}}
\caption{Antenna pattern in dBi of transmitter T located at $\rho_{\rm T}=0$.}
\label{fig:apatt}
\end{figure}

\section{Problem Statement}
\label{sec:mod}

We investigate a system consisting of a transmitter T, assumed to be static and non-strategic, that wants to establish a communication on a mmWave channel with a mobile receiver R, in an environment where an adversary A is present. The latter can be regarded as a mobile obstacle known to be placing itself between T and R to create blockage.

\subsection{Propagation scenario}
We limit the positions of T, R, and A onto a two-dimensional surface. We represent R and A's positions in polar coordinates $\mathbf{x} = (\rho, \vartheta)$ with T in the origin, i.e., $\rho_{\rm T} = 0$. (Note that $\vartheta_{\rm T}$ is undetermined, but this is irrelevant to the whole analysis).
The operating frequency of the channel is assumed to be $60$ GHz. 
The transmit antenna at T is modelled as the boresight beam of an $8 \times 4$ array of stacked patch antennas, placed at height of $1$ m. 
Its beam pattern, shown in Fig.\ \ref{fig:apatt}, has a $3$ dB half power beam width of $12.9^{\circ}$, a gain of $20$ dB and a side-lobe level of $-13.3$ dB.  Two sidelobes are present at $21^{\circ}$ and $38^{\circ}$, whereas nulls are present at $15^{\circ}$, $30^{\circ}$, and $50^{\circ}$. Due to the directivity of the patch antennas composing the array, beyond $60^{\circ}$ the antenna pattern decays smoothly and has very low amplitude. For this reason, we limit the range of angles considered in this analysis between $0^{\circ}$ and $60^{\circ}$.

The radiation pattern of the antenna would also extend to the interval $[-60^{\circ}, 0^{\circ}]$ but we exclude that region for the mobile players for symmetry considerations. Indeed, we are interested in understanding the mutual positioning of the players and the negative angular coordinates would convey the same meaning, but since they would correspond to a different statistical value of channel parameters, they would be a cause for higher deviations in the results.

\begin{figure}[tb]
\begin{center}
\resizebox{\figwb}{!}{\begin{tikzpicture}

\def\a{50}
\def\b{30}
\def\c{3}

\filldraw[color=black, fill=black!5, very thick] (0,0) circle (4pt) node[anchor=east, left = 2]{\footnotesize T};

\draw[black, thick, ->] (0,0) -- (3.5,0);
\draw[black, thick, ->] (0,0) -- (0,3);

\filldraw[color=red, fill=red!5, very thick]({1.5*cos(\a)}, {1.5*sin(\a)}) circle (0.4) node[above = 9]{\footnotesize A};
\draw[red, thick, ->] (0,0) -- ({1.5*cos(\a)}, {1.5*sin(\a)});
\draw [red,thick,domain=0:\a] plot ({0.8*cos(\x)}, {0.8*sin(\x)});
\draw (0.3,0.7) node[red] {\small $\rho_{\rm A}$};
\draw (1,0.3) node[red] {\small $\vartheta_{\rm A}$};

\filldraw[color=blue, fill=blue!5, very thick]({\c*cos(\b)}, {\c*sin(\b)})circle (4pt) node[anchor=west, right = 2]{\footnotesize R};
\draw[blue, thick, ->] (0,0) --  ({\c*cos(\b)}, {\c*sin(\b)});
\draw [blue,thick,domain=0:\b] plot ({2*cos(\x)}, {2*sin(\x)});
\draw (1.8,1.3) node[blue] {\small $\rho_{\rm R}$};
\draw (2.2,0.3) node[blue] {\small $\vartheta_{\rm R}$};

\draw [gray,domain=0:60] plot ({3*cos(\x)}, {3*sin(\x)});
\draw (3, -0.2) node[black] {\scriptsize $3\;$m};

\end{tikzpicture}}
\caption{Outline of the scenario. The angles $\vartheta_{\rm R}$ and $\vartheta_{\rm A}$ are aligned with the antenna pattern of the transmitter.}
\label{fig:scen}
\end{center}
\end{figure}
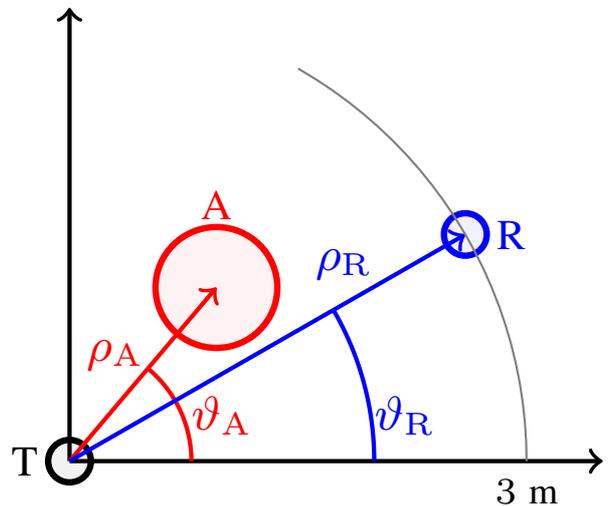

The receiver R is located at $\mathbf{x}_{\rm R} = (\rho_{\rm R}, \vartheta_{\rm R})$, where $\rho_{\rm R} = 3$ m. Thus, R is located at a fixed distance from the origin, but at a variable angle $\vartheta_{\rm R} \in [0^{\circ}, 60^{\circ}]$. Instead, the obstacle A is placed at a set of different distances between $1$ and $2.5$ m, namely $\rho_{\rm A} \in \{ 1+0.25k \}$ m, with $k=0, \dots, 6$, and the same options for the angular coordinate, i.e., $\vartheta_{\rm A} \in [0^{\circ}, 60^{\circ}]$. The physical shape of the obstacle A is a cylinder made of perfect electric conductor material, $1.75$ m tall, and of diameter equal to $0.5$ m.\footnote{This is actually meant to give a rough idea of a blockage caused by a person intentionally standing in between \cite{ghaddar2007conducting}.}
A pictorial representation of the positions of the involved elements is given in Fig.\ \ref{fig:scen}.

We remark that this represents a choice of a relatively small-scale propagation environment, which allows us to employ precise microwave simulation tools to compute the propagation scenario. Also, our main focus here is on the blockage-peeking interaction over the line of sight, which is present regardless of the distance. It would be clearly possible to extend many consideration to a larger scale propagation context (i.e., a bigger room) but this would likely require a statistical description of other obstacles possibly present in the area. Similarly, the choice of an adversary being an obstacle with circular symmetry is made to simply the reasonings about the obstructions caused. Both these issues, i.e., a statistical characterization of random obstacles along the path and the size of their chord intercepting the transmission beams in mmWave communications, are discussed in \cite{abdelnabi2020outage}, so the present analysis can be extended along these lines.

We assume a transmitter power of $10 \log_{10} (P_t) = 0$ dBm and an additive white Gaussian noise $n$ at the receiver with power $10 \log_{10}(P_{n}) = -100$ dBm. At the receiver's side, the antenna is also at height of $1$ m, and perfectly isotropic; this is meant for the sake of simplicity, since the whole directivity issues are now concentrated at the transmit antenna, and also because this gives a more interesting game theoretic perspective to the choice of locations that may have a better LoS but are aligned with secondary lobes of the transmit antenna pattern.
 
For the sake of this analysis, the channel gain can be decomposed into three main components. First of all, we consider an LoS component $r_1$, which is present whenever A is not between T and R. We also have a component $r_2$ scattered by A that reaches R. Finally, we include a multipath fading component $r_3$, coming from reflection of other objects present in the scenario, e.g., the ground and also walls for an indoor simulation.

The first two components $r_1$ and $r_2$ have been evaluated together, as a received complex term $r_{1,2}$, through Computer Simulation Technology microwave studio, 2022 edition \cite{cst}, which is a state-of-the-art platform for electromagnetic propagation simulation. These values are deterministically computed considering a free space propagation scenario and given all possible combinations of $\mathbf{x}_{\rm R}$ and $\mathbf{x}_{\rm A}$. The third component $r_3$ is superimposed as a randomly generated variable with Rayleigh distribution and independent of the positions of R and A. We assume that this component is, on average, equal to $-97$ dB. For each simulation, we generate $50$ different realizations of this component.

It is worth noting that, while the presence of A can block the LoS, it also causes more scattering. As a result, the power of the component $r_2$ also depends on the power distributed around by A; in certain situations, it might be that $r_2$ becomes the dominant component (i.e. $\vert r_2 \vert^2 > \vert r_1 + r_3 \vert^2$). This hints at that the best strategy for A is not necessarily blocking the main beam, which might cause $r_2$ to be large.

Given the channel gain and the transmitted power terms, it is immediate to compute the signal-to-noise ratio and, from this, the spectral efficiency of the system as
\begin{equation}
\nu = \log_2 \left( 1 + \frac{P_t \vert r_{1,2} + r_{3} \vert^2}{P_n} \right)
\label{eq:shannon}
\end{equation}

It can be remarked that in the absence of the obstacle causing blockage, the spectral efficiency experienced by R would be due to the LoS component with the maximum antenna gain. The latter is approximately corresponding to the free-space path gain of $-77.55$ dB plus $20$ dB of antenna gain, assuming a negligible impact of the $r_3$ component. As a result, the channel gain is $-57.55$ dB and, accordingly, Shannon's formula in (\ref{eq:shannon}) gives $\nu= 14.1$ b/s/Hz.

\subsection{Strategic interaction}

We apply this physical setup by considering R and A to be adversaries in a game, i.e., players with contrasting objectives, which is captured by the game being \emph{zero-sum} \cite{dasilva2011game,altman2007jamming}: the \emph{value} of the game can be regarded as the spectral efficiency, so that player R has utility $u_{\rm R} = \nu$, while A has utility $u_{\rm A} = -u_{\rm R} = -\nu$. 

Indeed, the fundamental modeling aspect of game theory is that individual moves of the players are combined together to determine an outcome that depends on the \emph{joint} strategic choices of all players. This is then translated again at the individual level of the players by computing their payoffs as player-specific utility functions \cite{tadelis}. The challenge of game theory is to identify individual choices that are desirable for the players based on their utilities, even though their goodness does not solely depend on an isolated strategic planning, but must also keep in consideration what the other player(s) do.

In our context, it is clear that these three elements of game-building are naturally present by virtue of the physical layer interaction. If we consider a scenario where players R and A choose their positions, so as to jointly determine the outcome as $({\bf x}_{\rm R}, {\bf x}_{\rm A})$, their individual choices are combined together into a resulting performance that depends on both. Even more so, if the players are driven by contrasting objectives, there is no inherently good or bad choice, since the goodness of a position clearly depends on what the other does. This is captured by the game theoretic concept of NE, which in a zero-sum setup like this can be obtained through minimax computation \cite{moon2015minimax}.

The key point is that both players R and A act \emph{strategically}, which makes our contribution differently from all the investigations where the receiver simply tries to avoid an unwanted blockage, since here this task is made harder by the malicious intention of A that is also able to react to R's countermoves.
We also remark that, while for numerical tractability we will consider a discrete search space for the nodes position, even a continuous version would in principle be solvable if ${\bf x}_{\rm R}$ and ${\bf x}_{\rm A}$ move within finite boundaries \cite{glicksberg20169}.

\section{Game Theoretic Analysis}
\label{sec:gam}
We frame the above problem as a static game of complete information, where R and A just choose, independently and unbeknownst to each other, a location within their own set of allowed positions.
Specifically, we consider the game in normal form to be represented by: (i) the set of players, R and A; (ii) their individual set of available actions, written as $\mathcal{P}_{\rm R}$ and $\mathcal{P}_{\rm A}$, respectively; and (iii) their individual utilities. The last point is, as discussed in the previous section, simply determined by setting the game as zero-sum, with a value equal to spectral efficiency, where R is the maximizer and A is the minimizer.

\begin{figure*}[t!h]
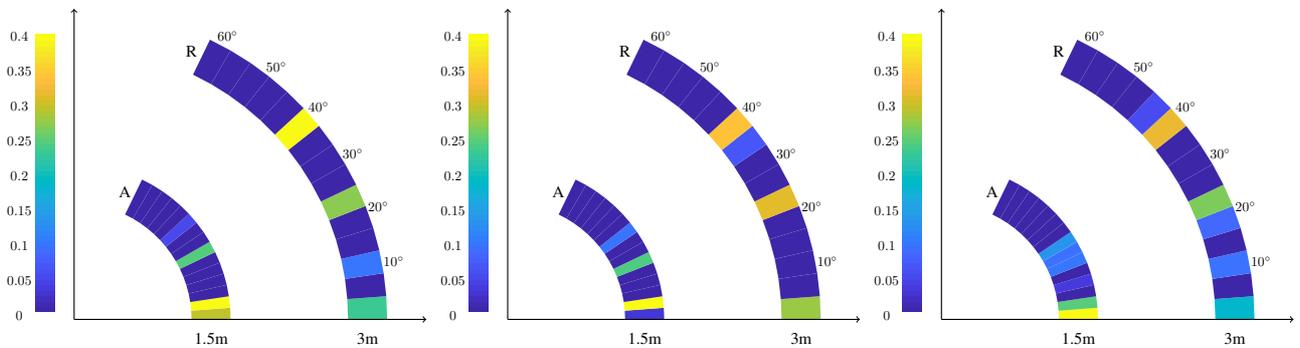

  \centering
         \resizebox{\figwc}{!}{\input{figures/heatmap_example}}
         \resizebox{\figwc}{!}{\input{figures/heatmap_example_5}}
         \resizebox{\figwc}{!}{\input{figures/heatmap_example_6}}
        \caption{Samples of mixed NE found.}\vspace{-0.1cm}
        \label{fig:sample}
\end{figure*}

As described in the scenario of Section \ref{sec:mod}, we consider pre-set values for $\rho_{\rm R}$ and $\rho_{\rm A}$, while the strategic choice involves $15$ possible options for both $\vartheta_{\rm R}$ and $\vartheta_{\rm A}$, as one of the values
\begin{equation}
\alpha_k = \frac{k}{7} \; 30^{\circ} \qquad \textrm{for $k=0,\dots,14$}
\label{eq:angular}
\end{equation}
Actually, this means that $\mathcal{P}_{\rm R} = \{ \vartheta_{\rm R}^{(k)} \}_k$ and $\mathcal{P}_{\rm A} \{ \vartheta_{\rm A}^{(k)} \}_k$ just involve the angular positions and are ultimately both equal to $\{ \alpha_k \}_k$, i.e., the only parameter of choice is the angular location chosen from the set for both and applied to a pre-set distance $\rho$ (that is however different for R and A, since A's is closer to T in order to cause blockage). This means that R and A can decide whether to locate themselves along the horizontal direction ($\alpha_0 = 0^{\circ}$) where the primary lobe of radiation from T is located, which would imply a stronger receiver signal in the absence of the obstacle, or to increase their angular position, up to $60^{\circ}$ which is the highest meaningful value of the radiation pattern from T. For numerical tractability, the whole interval from $\alpha_0 = 0^{\circ}$ to $\alpha_{14} = 60^{\circ}$ is divided into $15$ equally spaced angular positions. 

A key assumption for a game theoretic context would be that of \emph{complete information}, implying that whatever mentioned above, and in particular the existence of the obstacle itself, is common knowledge among the players \cite{tadelis}. This does not only mean that the players are aware of each other existence and parameters, but they are also fully able to derive rational conclusions based on such information. In particular, we point out that in our problem, the legitimate transmitter is fully informed that the adversary is present and trying to obstruct the communication, but the countermove to this can only be to peek around the obstruction. Clearly, catching games can also be analyzed (as we did in \cite{perin2021adversarial}) but they would be inherently different from the one under analysis here.
Another direction out of the scope of the present paper, but certainly possible as a future extension, would be a scenario with incomplete information about the \emph{possible but not certain} presence of a malicious obstacle, where, beyond finding a good path to the destination, the legitimate transmitter can also be interested in performing adversarial detection as well \cite{guglielmi2017analysis}.

The general approach to find the working point of a system modeled as a static game of complete information would be to derive its NE, which is the main solution tool for this case.
An NE can be seen as an outcome where players have no incentive for unilateral deviation \cite{dasilva2011game} and is therefore usually regarded as a stable working point, that rational players are able to anticipate and choose.
In all actuality, this would be true only when the NE is unique; however, this is fortunately true in the case under consideration.

The theoretical motivation for this result is to be sought in the representation of the game as zero-sum, and also in that this can be loosely seen as a discoordination game \cite{badia2019game}, where the different objectives of the players translate in that the maximizer wants them to play as differently as possible, whereas the minimizer desires their moves to be similar. In general, being zero-sum implies several interesting properties for the game \cite{scalabrin2015zero,moon2015minimax}, especially for the computational complexity of finding an NE, which boils down to finding a local critical point (a maximum or a minimum of the value depending on what respective player is considered). 
Also, if multiple NEs are present, they must all give identical values of the payoff. 

In our problem, given that the value is related to a continuous quantity (the spectral efficiency), this gives a theoretical guarantee of the uniqueness of the NE, which is also confirmed by numerical experiments. Yet, it is also likely that there are multiple working points for the system that get a value very similar to the one at NE, despite the actual positions of the players being different.
This justifies a purely numerical analysis averaged over different fading realizations, for which we expect that, in spite of the presence of even big local statistical variations, a general trend would emerge, which is investigated in the next section.


\section{Numerical Results}
\label{sec:res}

We discuss the findings gained from solving the static game of complete information, and how they can be expanded to more advanced scenarios. 

\subsection{Solution of the static game}

We consider R as placed at a distance $\rho_{\rm R}=3$ m from the transmitter, with $15$ available angular positions $\{ \vartheta_{\rm R}^{(k)} \}_k = \{ \alpha_k \}_k$ as per (\ref{eq:angular}).
Similarly, A has the same available angular positions $\{ \vartheta_{\rm A}^{(k)} \}_k = \{ \alpha_k \}_k$ but the distance from the transmitter, $\rho_{\rm A}$ is chosen as one of $7$ values in $[1.00:0.25:2.75]$ m. Notably, this is not part of the strategic choice, we are just considering different, yet pre-set, distance values for each set of evaluations.

The game in normal form can be represented through a matrix \cite{tadelis} and its solution via the NE is computed through the online solver available at \cite{avis2010enumeration}.
The typical result is a single mixed NE, for the reasons mentioned previously, whose support, i.e., the set of strategies played with non-zero probabilities, is a rather sparse choice of different angular positions, often involving non-adjacent values. Examples of NEs for the case where $\rho_{\rm A} = 1.5$ m are shown in Fig.\ \ref{fig:sample}. 

It can be seen that both players are likely to choose low angles, around $\alpha_0 = 0^{\circ}$, i.e., along the main beamforming lobe of the transmitter. However, this is not the most likely chosen position in many cases. For R, it is often convenient to go around the obstacle, usually selecting $\alpha_6$ and $\alpha_{10}$ (that turns out to be the most common choice), aligning the receiver apparatus on the secondary lobes and/or the best reception points of the specific fading realization, while at the same time evading A, and possibly benefitting from its scattering. As a strategic reaction, the adversary A also moves around trying to obstruct the reception, but to a more limited extent since it is still more convenient to block the main direction of transmission, so the main contribution to the support is usually given by actions $\alpha_0$ and $\alpha_1$, with the latter being actually slightly higher. This could actually look like a consequence of a border effect due to limiting the available angles to only positive values. However, side evaluations confirmed that even in the case negative angles are considered, the adversary still prefers to occupy a position that is a bit off the primary lobe, since the LoS will be obstructed anyways but also other positions can be covered.

Thus, it can be concluded that very often the NE corresponds to R selecting one of the main positions that allow to go around the obstacle; these positions are actually far apart and chosen with different probabilities. The actual choice of this ``mixture'' is to exclusively pick one of them, but the probabilities of each are determined so as to offer the least advantage to the adversary. Conversely, the adversary also randomizes but typically lingering around the main direction, and actually giving a higher probability to a tilted position so as to better cover the radiation pattern also when the receiver is located in the other positions.
This is an interesting conclusion that justifies the whole analysis and is only visible within a full-fledged characterization of the propagation scenario.

\begin{figure}[tb]
\begin{center}
\resizebox{\figw}{!}{\input{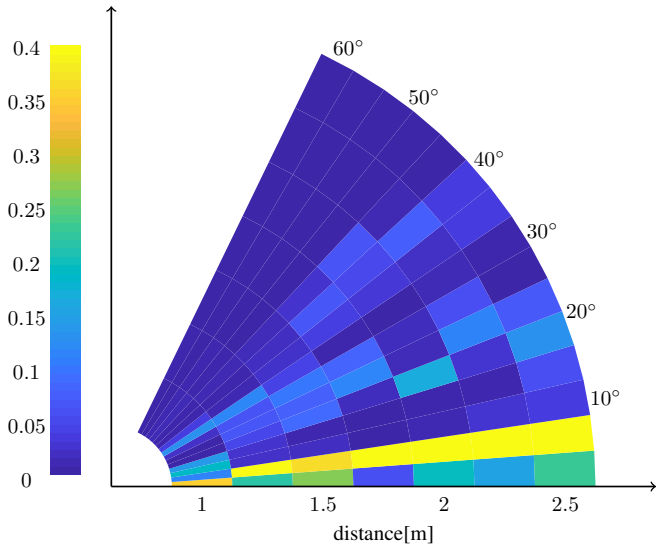}}
\caption{Average placement of the adversary at the NE}
\label{fig:posA}
\end{center}
\end{figure}

\begin{figure}[tb]
\begin{center}
\resizebox{\figw}{!}{\input{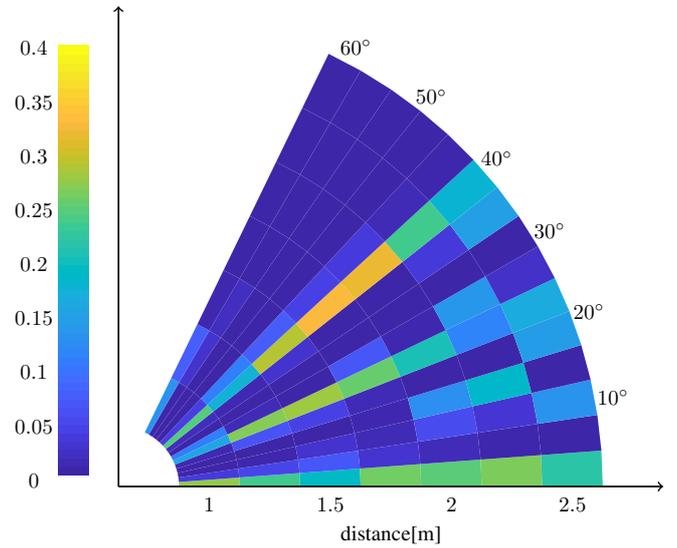}}\caption{Average placement of the receiver at the NE}
\label{fig:posR}
\end{center}
\end{figure}

To better understand the strategic choices of the players, it is convenient to average over multiple fading realizations, which obtains the results shown in Figs.\ \ref{fig:posA} and \ref{fig:posR} for A and R, respectively.
It is worth noting that, while Fig.\ \ref{fig:posA} reports an actual different position for the adversary according to the heatmap, in Fig.\ \ref{fig:posR} the value of $\rho_{\rm R}$ is always $3$ m instead, but the heatmap is kept with the same geometry for consistency.

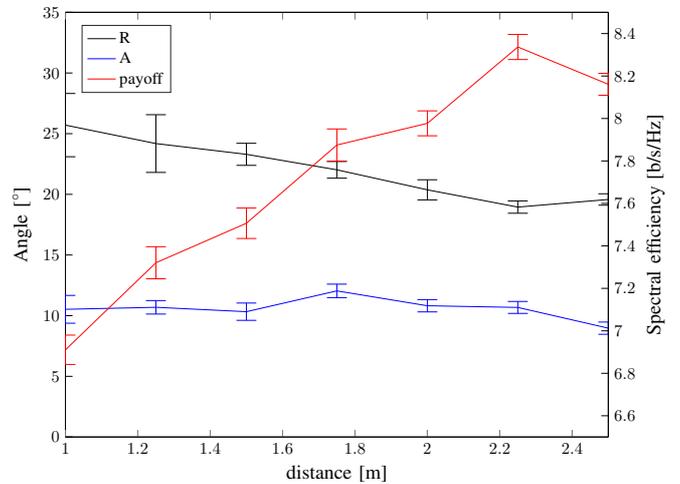
\begin{figure}[tb]
\begin{center}
         \resizebox{\figwd}{!}{
%
%
\begin{tikzpicture}

\begin{axis}[%
width=4.521in,
height=3.566in,
at={(0.758in,0.481in)},
scale only axis,
xmin=1,
xmax=2.5,
xticklabels=\empty,
separate axis lines,
every outer y axis line/.append style={black},
every y tick label/.append style={font=\color{black}},
every y tick/.append style={black},
ytick pos=left,
ymin=0,
ymax=35,
ylabel={$\text{Angle [}^\circ\text{]}$},
ylabel style={font=\large},
axis background/.style={fill=white},
legend style={legend cell align=left, align=left, draw=white!15!black}
]

\addplot [color=black, forget plot]
 plot [error bars/.cd, y dir=both, y explicit, error bar style={line width=0.5pt}, error mark options={line width=0.5pt, mark size=6.0pt, rotate=90}]
 table[row sep=crcr, y error plus index=2, y error minus index=3]{%
1.00	25.7055513428571	2.60715463806231	2.60715463806231\\
1.25	24.1852413428571	2.37860568805711	2.37860568805711\\
1.50	23.2979226857143	0.90939980589158	0.90939980589158\\
1.75	22.0208835428571	0.687026700383285	0.687026700383285\\
2.00	20.3716460571429	0.830035414004814	0.830035414004814\\
2.25	18.9491064	0.506916511702091	0.506916511702091\\
2.50	19.5757040571429	0.470569105038372	0.470569105038372\\
};\label{plotRX}
\addplot [color=blue, forget plot]
 plot [error bars/.cd, y dir=both, y explicit, error bar style={line width=0.5pt}, error mark options={line width=0.5pt, mark size=6.0pt, rotate=90}]
 table[row sep=crcr, y error plus index=2, y error minus index=3]{%
1.00	10.5236512285714	1.14502039881465	1.14502039881465\\
1.25	10.6890931714286	0.555347703964837	0.555347703964837\\
1.50	10.3277674285714	0.712554971395423	0.712554971395423\\
1.75	12.0420367714286	0.556009386467921	0.556009386467921\\
2.00	10.8170262857143	0.506350853599783	0.506350853599783\\
2.25	10.6790797714286	0.493401637539871	0.493401637539871\\
2.50	8.97008545714286	0.503868025671463	0.503868025671463\\
};\label{plotH}

\end{axis}

\begin{axis}[%
width=4.521in,
height=3.566in,
at={(0.758in,0.481in)},
scale only axis,
xmin=1.00,
xmax=2.50,
xlabel style={font=\large},
xlabel={distance [m]},
separate axis lines,
every outer y axis line/.append style={black},
every y tick label/.append style={font=\color{black}},
every y tick/.append style={black},
ymin=6.5,
ymax=8.5,
ylabel={Spectral efficiency [b/s/Hz]},
ylabel style={font=\large},
axis y line*=right,
legend style={legend cell align=left, align=left, draw=white!15!black,at={(0.03,0.97)},anchor=north west}
]

\addplot [color=red, forget plot]
 plot [error bars/.cd, y dir=both, y explicit, error bar style={line width=0.5pt}, error mark options={line width=0.5pt, mark size=6.0pt, rotate=90}]
 table[row sep=crcr, y error plus index=2, y error minus index=3]{%
1.00	6.9107773686774	0.069102323931836	0.069102323931836\\
1.25	7.3206418047374	0.0755107832270019	0.0755107832270019\\
1.50	7.5065547130008	0.0717139731733139	0.0717139731733139\\
1.75	7.8751259974994	0.0756747773636394	0.0756747773636394\\
2.00	7.9772569154822	0.0587135469364427	0.0587135469364427\\
2.25	8.3367546658474	0.058951038461363	0.058951038461363\\
2.50	8.160999765287	0.051732527840516	0.051732527840516\\
};\label{plotPO}

\addlegendimage{/pgfplots/refstyle=plotRX}\addlegendentry{R}
\addlegendimage{/pgfplots/refstyle=plotH}\addlegendentry{A}
\addlegendimage{/pgfplots/refstyle=plotPO}\addlegendentry{payoff}

\end{axis}
\end{tikzpicture}%
}
\caption{Summary of relevant metrics. Confidence intervals are at $3$ standard deviations for $50$ simulation runs.}
\label{fig:posM}
\end{center}
\end{figure}

Once again, it is highlighted that the NE, on average, corresponds to R choosing the three aforementioned angular positions with similar probabilities, while A is usually located between the first two positions but preferring $\alpha_1$ over $\alpha_0$. However, there are also some local variations depending on fading, and border effects.
In particular, the closer the adversary to the transmitter, the less likely the choice of $\vartheta_{\rm A} = \alpha_1$.
This seems to imply that A chooses this value to better block the lateral positions, but when the obstacle is placed closer such a blockage is already strong enough, so it becomes more convenient to choose $\alpha_0$. Conversely, the choices of $\vartheta_{\rm R}$ as equal to $\alpha_6$ or $\alpha_{10}$ are much more likely when the obstacle is placed at intermediate values rather than very close to the transmitter (in which case its disturbance is very big anyways) or very far from it. 

In the latter case, we observe a change in the general trend, in that the obstacle may actually place itself in front of the receiver and therefore the best strategy of R changes even more, in a counter-intuitive way. We see that at the same time the likelihood of $\alpha_0$ is increased, but also $\alpha_6$ and $\alpha_{10}$ are no longer preponderant. This seems to imply that, beyond the direct positioning aligned with the main beam, the receiver also randomizes more on the other positions. We remark that when the obstacle is far from the transmitter, even a minimal change of position is enough to avoid the blockage.

To give a quantitative summary of these findings, we show in Fig.\ \ref{fig:posM} the average positions of R and A, as well as the resulting value of the game, as functions of $\vartheta_{\rm A}$. For what concerns the positions, some caution should be exerted in that this graph plots the average of an average, that is, the stochastic average over multiple fading realizations of the average position at the NE, which is itself probabilistic.
This is just done for the sake of graphical representation, and because the scale of the simulation setup is sufficiently large, but it is imperative not to confuse averaging statistical realizations with choosing intermediate positions. In other words, choosing two positions, having some separation in between, with 50/50 probability, is clearly different than choosing to stay in the middle between them (a choice that is actually never made).
With all these caveats in mind, the figure seems to imply that the overall average position of the obstacle is basically unchanged even at different distances, since the corresponding curve is very flat. On the other hand, the angle of the receiver position becomes narrower as A approaches, with the exception of the last point in the curve, implying that this trend is reversed when the obstacle becomes very close.

It is also interesting to see the average value of the payoff obtained at the NE, as a function of $\rho_{\rm A}$. This is also reported in Fig.\ \ref{fig:posM}. Since from (\ref{eq:shannon}), we computed a maximum spectral efficiency of $14.1$ b/s/Hz in the absence of the obstacle, we see that the presence of A causes to lose almost half of that value, depending on the position. The figure also implies that the closer the obstacle to the transmitter, the harder to go around it. This also shows that, despite the alignment for different choices of the $\vartheta$s being the same, there is also a non-negligible effect of absolute distance on mmWave propagation, possibly enhanced by the strategic interaction. 
About this last point, we can also remark that if the receiver was aligned on the main secondary lobe, but without any obstacle, we can get through analogous computations that $\nu = 9.7$ b/s/Hz, which shows that the adversary is able to decrease this value as well, thanks to a strategic choice of blockage that prevents the receiver to be permanently at that location.

All of these reasonings show that the strategic interaction, even for a simple static game, is far from trivial. It would be interesting to extend the analysis to see what happens if both the receiver and the adversary have more degrees of freedom for the movement in the considered area, or the scenario is complicated by other propagation effects and transmission options.

\subsection{Discussion and extension to dynamic setups}

The general solution of the static game involves a mixed NE, which translates in a linear combination of actions. This needs to be understood on several different levels.
From the perspective of a strategic interaction, the mixed NE is the logical consequence of the adversarial setup \cite{moon2015minimax}. Ideally, this corresponds to the players attempting to leave the opponent in a state of indeterminacy about their actions, according to the indifference theorem \cite{tadelis}. Indeed, if either player was choosing a pure strategy, the counteraction would be easy to identify as a best response, and therefore it is convenient to randomize in order to be as little revealing as possible.

Given the evident connection between the radiation pattern of the transmitter and the strategic choices of the players, it is also tempting to read the NE as a limit choice to which the players tend. However, this would be inappropriate, since the NE is an inherently static concept, and any interpretation as the result of subsequent approximations is not right. This is especially relevant since the best choices of the receiver often combine non-adjacent angular positions, and it is more logical to think of the movements of the involved players as having some smooth transitions. This might imply that the equilibrium in a dynamic game is somehow different from the static one, giving a better upper hand to the adversary, if the receiver cannot change location at will but is limited to adjacent positions. Also, the search space would increase quite rapidly and implies that meta-heuristic search tools would be needed \cite{perin2021reinforcement}.

\section{Conclusions and Future Work}
\label{sec:concs}

We presented a novel adversarial game for mmWave communications, involving a player that tries to obstruct the LoS of a receiver, who can in turn react strategically. We called it a ``Blockage-Peeking'' game, which in the present paper is formalized as a static game of complete information.
We described the resulting NE through both theoretical and numerical evaluations, showing how it correlates to the radiation pattern of the transmitter.
We also discussed how this can be extended to a dynamic setup, which would require more advanced solution tools and can be the scope of future investigations \cite{etesami2019dynamic}.

An immediate extension of this investigation concerns the physical setup, since we placed the nodes in a relatively small propagation environment and also considering other features of mmWave communications, such as ultra-fast beamsteering \cite{bonjour2015ultra}. Our intention here was not to claim generality of the analyzed scenarios, but rather to consider a strategic feature that is rarely addressed in the literature. Possibly, this identifies an interesting direction for future research in combination with other features of mmWave scenarios.

Even though the relative distances between transmitter, receiver, and obstacle were varied in the analysis, but just as parameters in the game, another extension of the present investigation would be to take them as strategic choices for the players.
As hinted in the discussion of the results, it would be interesting to also permit the nodes to change their distances on a strategic basis, as an evasion move for the receiver, for example. 
More in general, further extensions of the study may involve a wider mobility scenario, where movement is allowed along other coordinates as well. 

Finally, the arrangement as a game of complete information requires the assumption that all the propagation parameters are known to the players, which is a clearly optimistic assumption, even though it is not unrealistic to assume that they can be estimated somehow. Still, more advanced game theoretic investigations are also possible, where Bayesian games are employed, to keep into account that the players may have incomplete (i.e., probabilistic) information \cite{quer2013inter}. This can be true for both the channel parameters, that are inherently stochastic, or the presence itself of the adversary that may be uncertain.
 

\section*{Acknowledgment}
This project has received funding from the European Union's Framework Programme for Research and Innovation Horizon 2020 under Grant Agreement No.\ 861222 (EU H2020 MSCA MINTS).
 
\IEEEtriggeratref{22}
\bibliographystyle{IEEEtran}
\bibliography{IEEEabrv, jambad}
\end{document}